\documentclass[twoside,onecolumn]{article}

\usepackage{blindtext} 

\usepackage[sc]{mathpazo} 
\usepackage[T1]{fontenc} 
\usepackage{microtype} 

\usepackage[english]{babel} 

\usepackage[hmarginratio=1:1,top=32mm,columnsep=20pt]{geometry} 
\usepackage[hang, small,labelfont=bf,up,textfont=it,up]{caption} 
\usepackage{booktabs} 

\usepackage{lettrine} 

\usepackage{enumitem} 
\setlist[itemize]{noitemsep} 

\usepackage{abstract} 

\usepackage{titlesec} 
\renewcommand\thesection{\Roman{section}} 
\renewcommand\thesubsection{\roman{subsection}} 
\titleformat{\section}[block]{\large\scshape\centering}{\thesection.}{1em}{} 
\titleformat{\subsection}[block]{\large}{\thesubsection.}{1em}{} 

\usepackage{fancyhdr} 
\usepackage{titling} 

\usepackage{hyperref} 

\pretitle{\begin{center}\Huge\bfseries} 
\posttitle{\end{center}} 
\title{Square Kilometre Array Science Data Challenge 1} 
\author{%
\textsc{Anna Bonaldi \& Robert Braun, for the SKAO Science Team} \thanks{SKASDC1@skatelescope.org} \\[1ex] 
\normalsize SKA Organization, Jodrell Bank, Lower Withington, Macclesfield, \\ \normalsize Cheshire, SK11 9DL, United Kingdom \\ 
}
\date{\today} 
\renewcommand{\maketitlehookd}{%
\begin{abstract}
\noindent 
The Square Kilometre Array (SKA, https://skatelescope.org) will be the world's largest radio telescope. SKA Science Data Challenges will be regularly issued to the community as part of the science preparatory activities. The purpose of these challenges is to inform the development of the data reduction workflows, to allow the science community to get familiar with the standard products the SKA will deliver, and optimise their analyses to extract science from them. These challenges may consist of real data from currently operating radio facilities or of simulated SKA data. The purpose of this document is to provide information on how the SKA Science data challenge \#1 (SDC1) has been produced and to set the challenge for the community. For more information on how to take part in the challenge and to download the data see https://astronomers.skatelescope.org/ska-science-data-challenge-1/ 
\end{abstract}
}

\begin{document}

\maketitle


\section{Introduction}


For the purpose of the data challenges, SKA data at different stages along a data reduction workflow have been broadly categorised into four main Data Layers (DL)
\begin{itemize}
\item	DL1: Raw data. These are typically observations of a few hour duration, consistent with a single SKA scheduling block. Data are typically uncalibrated and the main focus of a challenge might be to inform the calibration strategy, its implementation, efficiency and scalability.
\item	DL2: Observatory Data Products. These are what the Science Data Processor (SDP) will typically provide: based on calibrated data, from a list of standard products (see Table 1), corresponding to a few hour duration observation, consistent with a single SKA scheduling block. The focus of a challenge at this DL is to carry out the kind of processing that Science Regional centres (SRCs) will perform to provide the Principal Investigator/Key Science Project teams with added value data products.
\item	DL3: Advanced Data Products. These are the typical SRC output: products based on multiple SKA scheduling blocks (e.g. images derived from SKA data over a wide area and/or a deep integration). The objective of data challenges at this layer is to extract science from the data, with a focus on algorithm development.
\item DL4: Scientific results. This is a proposal-specific product, that is ultimately the goal of the whole observation and analysis. 
\end{itemize}

Data for SKA Science Data Challenges can be made available at any of the four stages. The objective for each data challenge is defined, however usage of the data for other purposes is encouraged. Usage of the data beyond the defined challenge should be suitably acknowledged. This dataset can be referenced as SKAO data challenges, Science Data Challenge \#1; other references can be found at the end of this document.  

\begin{table*}
\caption{List of standard SDP products in the current design documenation}
\centering
\begin{tabular}{l}
\toprule
Local Sky Model (Developed for calibration)  \\
Calibrated Visibilities (Typically with substantial time and frequency averaging) \\
Image Products: UVgrids (Gridded visibilities and data weights) \\
Image Products: Image cubes  \\
Sieved Pulsar and Transient Candidates\\
Pulsar Timing Solutions\\
Dynamic spectrum data\\
Transient Buffer data\\
Transient Source Catalogue\\
Science Alert Catalogue\\
Science Product Catalogue\\
\bottomrule
\end{tabular}
\end{table*}


\section{SKA Science Data Challenge \#1 dataset description}
The SKA Science Data Challenge \#1 (SDC1) release consists of 9 image files, in FITS format. Each file is a simulated SKA continuum image in total intensity at 3 frequencies:
\begin{enumerate}
\item	560 MHz, representative of SKA Mid Band 1
\item	1.4 GHz, representative of SKA Mid Band 2
\item	9.2 GHz, representative of SKA Mid Band 5
Furthermore, 3 telescope integration depths per frequency are provided: 
\begin{enumerate}
\item	8 h, representative of a single observation (DL2/3)
\item	100 h, representative of a medium-depth integration (DL3)
\item	1000 h, representative of a deep integration (DL3).
\end{enumerate}
\end{enumerate}
The Field of View (FoV) was chosen for each frequency to contain the primary beam for a single telescope pointing out to the first null. This gives a map size of 5.5, 2.2 and 0.33 degrees on a side for 560 MHz, 1.4 GHz and 9.2 GHz respectively. The number of pixels on a side is always 32,768, which gives a pixel size of 0.60, 0.24, and 0.037 arcsec for 560 MHz, 1.4 GHz and 9.2 GHz, respectively.

The simulated field is nominally centred at RA=0, Dec=-30 for each map. The sky model is a plausible realization of the radio sky at those frequencies, but \emph{there is no attempt to make it similar to the actual sky at those coordinates}. The nine maps share the same sky model realizations, to allow cross-matching between frequencies and direct comparisons between results for different noise levels. 

The dataset is also accompanied by Ancillary Data:
\begin{itemize}
\item	The Primary Beam and Synthesized Beam for each frequency band are provided as FITS format images
\end{itemize}
And a training set:
\begin{itemize}
\item	A "Truth Catalogue" for each frequency band, listing all of the embedded sources and their properties within 5\% of the FoV area, as an ASCII format table. This Catalogue also serves as a submission template for the Data Challenge. 
\end{itemize}


\section{Sky model description}
The list of objects that populate the simulated fields was generated with the T-RECS simulation code (RD1), yielding catalogues of AGNs and SFGs, with integrated flux densities, sky coordinates, and size and shape information. 

Sources from these catalogues were injected into the simulated fields either as:
\begin{itemize}
\item	an extended source, if its major axis size (defined below) is larger than 3 pixels;
\item	a compact source, if its size is smaller than 3 pixels.
\end{itemize}
Given that the pixel size depends on frequency, the same source can be treated as extended in one map and as compact in another. Images of extended sources were generated according to their morphological parameters and then added as "postage stamps" to an image of the full field. Compact sources were added to the images with a Gaussian convolving kernel (described in more detail in Sec. 4).

The morphological model for the SFGs is an exponential Sersic profile, projected into an ellipsoid with a given axis ratio and position angle. The SFG images for the full field of view were generated with the Galsim-based pipeline developed for the SuperCLASS project (RD2). 

All AGN populations are treated as the same object type viewed from a different angle, so that steep-spectrum AGNs exhibit the typical double-lobe FRI/FRII morphology, while flat-spectrum ones exhibit a compact core component together with a single lobe viewed end-on. For the steep-spectrum AGNs we used a library of real, high-resolution AGN images (RD3), scaled in total intensity and size, and randomly rotated and reflected, to generate the postage stamps. A correction to the flux of the core (modelled for this purpose as a Gaussian having the FWHM of the original library image) has been applied in order to give it a flat spectral index. This means that the same AGN at different frequencies has a different core to lobes fraction. All flat-spectrum AGNs were added as a pair of components; one of milliarcsecond scale with a given "core fraction" of the total flux density and another of a specified larger size.  

The "size" of all AGN populations is defined as the Largest Angular Size (LAS) of the source.  The "size" of the SFGs is defined as the exponential scale-length of the disk.

\section{Telescope simulation description}
\subsection{The Gridded Sky Model}
The extended source images (with units of Jy/Pixel) were given a mild Gaussian convolution using a FWHM ($g$) of two pixels (yielding an image in units of Jy/"Gaussian smoothing Beam").

The three catalogues (SFGs, steep spectrum AGN and flat spectrum AGN) of compact objects were added to the image as elliptical Gaussian components. 

The compact SFGs are described by an integrated flux density, $I$, a major axis and minor axis size $(s_1,s_2)$ (with "size" defined as the exponential scale length) together with a major axis position angle,$\phi$, giving $(I,s_1,s_2,\phi)$. These sources are approximated by an elliptical Gaussian with FWHM $(w_1,w_2)$ in each dimension equal to $2^{1/2}$ times the exponential scale length in that direction, (ie. $w_1 = 2^{1/2} s_1$ and $w_2 = 2^{1/2} s_2$) yielding an intrinsic characterisation of each galaxy by $(I,w_1,w_2,\phi)$. 

Compact steep spectrum AGN are described by an integrated flux density, $I$, a major axis and minor axis size $(a_1,a_2)$ (with "size" defined here as the LAS of the source) together with a major axis position angle ,$\phi$, so $(I,a_1,a_2,\phi)$. These sources are approximated by an elliptical Gaussian with FWHM $(w_1,w_2)$ in each dimension equal to half the LAS in that direction, (ie. $w_1 = a_1/2$ and $w_2 = a_2/2$), yielding the characterisation, $(I,w_1,w_2,\phi)$.

Compact flat spectrum AGN are described by an integrated flux density, $I$, a size $f$ (with "size" defined here as the largest angular size (LAS)) together with a core fraction, $r_c$. These sources are approximated by a pair of circular Gaussians; a compact core with flux density $I_c = I \times r_c$, and Gaussian FWHM size of $w_c = g/10$ and a more extended end-on lobe with flux density $I_d = I  \times (1 - r_c)$, and Gaussian FWHM size of $w_d = f/2$, so a characterisation as $(I_c,w_c)$ and $(I_d,w_d)$.

For the purpose of gridding, the intrinsic source dimensions described above are added in quadrature to a two pixel FWHM gridding kernel, $g$, to give the gridded representation of each source $(I,b_1,b_2, \phi)$ where $b_1 = (w_1^2 + g^2)^{1/2}$  and $b_2 = (w_2^2 + g^2)^{1/2}$ . The image representation is normalised to be in units of Jy/Beam, with the "beam" FWHM $= g$. The complete gridded sky model is the sum of the extended source image and the compact source image.

\subsection{Application of the Primary Beam Response}

A simulation of the SKA primary beam response at a frequency of 1.4 GHz has been provided by the Dish Design Consortium (RD4). The average response of the two perpendicular linear polarisations was formed for a fixed telescope orientation.  In fact, the two individual linear polarisations are intrinsically elliptical with an axis ratio of about 1.09 prior to averaging. The gridded sky model has simply been multiplied with this static model of the primary beam. This is a significant simplification, since in practice the relative orientation of the beam pattern and sky would rotate during the course of an actual observation. The primary beam response at 560 MHz and 9.2 GHz was approximated by a rescaling of pixel size by the inverse frequency ratio. The three beam models are available for download and can be used for a primary beam correction of the apparent flux densities within the final images.

\subsection{Generation of Visibility Coverage, PSF and Noise Images}
Visibility data files were generated using the SKA1-Mid configuration (RD5) both with and without the inclusion of the 64 MeerKAT dishes. In the first case, 197 antenna locations are specified for use at 1400 MHz, while in the latter, 133 antenna locations are used (for 560 and 9200 MHz where it is assumed that the MeerKAT dishes are not equipped with feeds). The visibility sampling is based on 91 spectral channels that span a 30\% fractional bandwidth. For the three centre frequencies of (560, 1400 and 9200) MHz this meant (168, 420 and 2760) MHz of total bandwidth. The time sampling spanned -4h to 4h of Local Sidereal Time with an increment of 30s at an assumed Declination of -30. This sampling resulted in data files of about 8 GB size. This is much smaller than actual data files which would require sampling at a fractional bandwidth of 10$^{-5}$ and 0.14s to avoid significant smearing effects. Actual data files at this native resolution would be some 570 TB in size. 

Noise images and dirty point spread functions were generated from the visibility files. Gridding weights for the visibility data were determined as follows. All of the visibility samples were first accumulated in the visibility grid with their natural weights. This visibility density grid was subjected to an FFT-based convolution using a Gaussian convolving function with FWHM of 178m, corresponding to an image plane FWHM of (538, 215 and 33) arcsec at the three centre frequencies being simulated. This convolving function width was determined by trial and error to provide the best match to the sampling provided by the array configuration. The convolved natural density distribution was used to form so-called "uniform" weights for the visibilities (from the inverse of the local smoothed data density), after which a Gaussian taper was applied to achieve the most Gaussian possible dirty beam with a target FWHM of (1.5, 0.60 and 0.0913) arcsec at the (560, 1400 and 9200) MHz. The actual dirty beam dimensions were closely matched to the target specification. Dirty beam sidelobes for both the 133 and 197 dish configurations have maximum absolute values of $4 \times10^{-4} $and an RMS sidelobe level within the central $10 \times 10$ beam areas of $4 \times10^{-5}$. The image noise in these images is degraded from the naturally weighted image noise for the array by a factor of two. The noise images were rescaled in amplitude to represent an appropriate RMS fluctuation level for 8, 100 and 1000 hours of effective integration time.

\subsection{Simulated Deconvolution and Restoration of the Sky Model}

The data reduction workflow in the SDP is such that calibration and imaging happens on single scheduling block instances, generating Observatory Data Products that are accumulated and co-added in SRCs to produce deeper images. To reflect this, deconvolution has been simulated for all our images down to a depth appropriate for an 8 h observation. 

Specifically, the primary beam tapered sky model (as described in Sec 6.2 above) was clipped at a brightness of three times the expected RMS noise level for an 8 hour observation, namely (8.6, 2.1 and 1.3) $\mu$Jy/Beam at (560, 1400 and 9200) MHz. (Please see RD7 for a discussion of the anticipated imaging sensitivity.) The excess emission above this clipping level was subjected to a convolution to a final effective restoring beam size of (1.5, 0.60 and 0.0913) arcsec FWHM, while taking explicit account of the gridding function dimensions (a two pixel FWHM Gaussian that was subtracted away in quadrature). The residual brightness distribution, below the clipping level, was first subjected to a linear deconvolution of the Gaussian gridding function (with FWHM=g) and then convolved with the dirty PSF. The convolved peak and convolved residual images were then summed.

\subsection{Simulated Single Track and Deep Co-added Images}
The final data products were generated by adding the various noise images to the convolved sky model. RMS fluctuation levels in the final images are dominated by the noise images at the depths appropriate for 8 and 100 hours, but are elevated at the 1000 hour depth due to a contribution from residual sidelobe fluctuations resulting from the finite "deconvolution" depth.

\section{Limitations of the Simulated Data Products}
It's important to emphasize the known limitations of the current simulated data products, which in general lead to an unrealistically optimistic image quality. The level of realism will gradually increase with future simulations. 
\begin{enumerate}
\item	No calibration errors of any kind have been introduced into the simulated images. In practice instrumental, ionospheric and atmospheric gain fluctuations will all be present in an observation and these will be calibrated out to some precision using the strategy outlined in RD6. 
\item	No pointing errors are included in the simulation, nor has any account been taken of the time varying parallactic angle of the dish response relative to the sky. Both of these effects would be present in an actual observation and would need to be calibrated out.
\item	Only the average of the two linear polarisation beam patterns is used to approximate the telescope response (and this is static rather than allowed to rotate). Furthermore, the beam patterns are only simulated ones for the 15m diameter SKA dish design at a single reference frequency and are assumed to be identical for all antennas in the array. The simulated pattern is linearly scaled from the reference frequency to other frequencies of interest. Actual beam patterns will need to be determined once deployed and may well display differences amongst dishes. One obvious and significant issue is the very different primary beam pattern of the 64 MeerKAT dishes (of 13.5m diameter) relative to the SKA dishes. 
\item	Although each component of the sky model has a fully specified spectral energy density distribution, it is represented in the simulations with a fixed flux density at each of the three reference frequencies rather than with a flux density that varies within each observed band. 
\item	The imaging process has been quite crudely approximated. "Perfect" deconvolution is assumed to have been undertaken down to a depth of 3 times the thermal RMS noise level of an 8 hour observation. The residuals of this perfect deconvolution are then degraded with an approximate representation of the dirty PSF. However, even this PSF degradation is carried out using a convolution with a position invariant PSF, despite the field of view being extremely large in relation to the PSF. If the sky model had actually been incorporated into the visibilities, then it would have been necessary to achieve visibility sampling at a fractional bandwidth of 10$^{-5}$ and 0.14s as well as taking explicit account of the "w" coordinate of the baselines in order to avoid significant smearing effects during imaging and deconvolution. 
\end{enumerate}
\section{	The Challenge Defined}
The SKA community is invited to retrieve the SDC1 images and undertake:
\begin{enumerate}
\item	Source finding (RA, Dec) to locate the centroids and where appropriate the core positions
\item	Source property characterisation:
\begin{enumerate}
\item	integrated primary-beam corrected flux density (the primary beams are provided)
\item	core fraction (it is different from zero only for AGN)
\item	major and minor axis size, where size is one of (largest angular size, Gaussian FWHM, or exponential scale length). Conversions will be applied between the three sizes to always allow comparison with the input catalogue.
\item	major axis position angle, measured clockwise from due west.  
\end{enumerate}
\item	Source population identification (one of AGN-steep, AGN-flat, SFG)
\end{enumerate}
Results can be submitted using the standard SDC1 response template (defined by the Training Catalogue described in Section 4) and these will be compared with and scored against their agreement with the input catalogue. Only those submissions that adhere strictly to the formatting of the template will be processed. The accuracy of submitted results will be graded on:
\begin{enumerate}
\item	Reliability and completeness of sources found 
\item	Accuracy of property characterisation
\item	Accuracy of population identification 
\item	Overall response score based on the total number of real sources (less any false detections) found in the three 1000h images (the sum over the three frequency bands, while excluding the Training Area for each band defined in Section 4) multiplied by the fractional accuracy of the property characterisation and the population identification
\end{enumerate}
Participating teams are encouraged to lodge their interest with the authors of this description, by sending an email to \emph{skaSDC1@skatelescope.org}, so that they can be directly informed of any updates. A provisional deadline for submissions is 15 March 2019.  

The overall SDC1 winner will have the highest total score. Only the submission clearly identified as "final" by each participating team will be graded.


\end{document}